\documentstyle[12pt]{article} % Paper without sections(PhysLet)

\input psfig

\newcommand{\ds}{D^{\ast}}

\newcommand{\D}{D_{s}}
\newcommand{\Ds}{D^{\ast}_{s}}
\newcommand{\btablec}{\begin{table}[tbp] \begin{center}}
\newcommand{\etablec}{\end{center} \end{table}}
\newcommand{\intereqnvspace}{\vspace{-1.4ex}}
\newcommand{\beqn}{\begin{equation}}
\newcommand{\eeqn}{\end{equation}}

\newcommand{\be}{\begin{equation}}
\newcommand{\ee}{\end{equation}}
\newcommand{\bea}{\begin{eqnarray}}
\newcommand{\eea}{\end{eqnarray}}
\newcommand{\bean}{\begin{eqnarray*}}
\newcommand{\eean}{\end{eqnarray*}}

\newcommand{\gapproxeq}{\lower.7ex\hbox{$\;\stackrel{\textstyle>}{\sim}
\;$}}
\newcommand{\lapproxeq}{\lower .7ex\hbox{$\;\stackrel{\textstyle
<}{\sim}\;$}}
\newcommand{\pbibitem}[1]{\bibitem{#1}}
\newcommand{\plabel}[1]{\label{#1}}

\def\thefiglist#1{\section*{Figure Captions\markboth
{FIGURE CAPTIONS} {FIGURE CAPTIONS}}\list
{Figure \arabic{enumi}}
{\settowidth\labelwidth{Figure #1.}\leftmargin\labelwidth
\advance\leftmargin\labelsep
\usecounter{enumi}.}
\def\newblock{\hskip .11em plus .33em minus -.07em}
\sloppy}

\begin{document}
\begin{titlepage}
\begin{flushright} hep-ph/9507407 \\ RAL-95-035 \\ OUTP-95-13P \\
\end{flushright}
\vskip 2cm
\begin{center}
{\bf\large DO $\psi$ (4040), $\psi$ (4160) SIGNAL HYBRID
CHARMONIUM?\footnote{Work supported in part by the European Community
Human Mobility program ``Eurodafne'' contract CHRX-CT92-0026 and the
University of Cape Town.}}
\vskip 1cm
 {\bf F E Close}\\
{\it Rutherford Appleton Laboratory\\
Chilton, Didcot, Oxon
OX11 OQX,  U.K.\\ \rm fec@v2.rl.ac.uk}\\
{\it and} \\ {\bf P R Page}\\{\it
Theoretical Physics\\
1 Keble Road, Oxford, OX1 3NP, U.K.\\\rm p.page@ph.ox.ac.uk} \\
\end{center}
\vspace*{3cm}
\begin{abstract}
We suggest that $\psi$ (4040) and $\psi$ (4160) are
strong mixtures of ground
state hybrid charmonium at $\sim 4.1$ GeV
and the $\psi (3S)$ of conventional charmonium.
The $\Gamma^{e^+e^-}$, masses and total widths of the $\psi(4040)$
and $\psi(4160)$ are in accord with this
hypothesis.  Their hadronic decays are predicted to be
dominated by the $\psi (3S)$ component
and hence are correlated.
In particular we find a spin counting
relation $\Gamma (4160 \rightarrow D_sD_s^*) \sim 4 \Gamma (4040 \rightarrow
D_sD_s)$ due to their common $\psi(3S)$ component. For $D$ and $D^*$
production, using $\psi(4040)$ branching ratios as input, we predict that
the decay pattern of the $\psi(4160)$ will be very different
from that of the $\psi(4040)$. These predictions may be tested in
historical data from SPEAR, BES or at future Tau-Charm
Factories.
\end{abstract}
\end{titlepage}

The QCD sector of the Standard Model cannot be regarded as finally
established while the question of the existence of glueballs and hybrids
(hadrons where the gluonic modes are excited in the presence of quarks)
remains unresolved. This is now becoming rather critical. On the theory
side a consensus on spectroscopy and dynamics is emerging from lattice QCD
and model simulations thereof whereas on the experimental side,
the increasing
concentration of resources at the high energy frontier threatens to
leave a hole in this essential area of the Standard Model.

We have initiated a programme to evaluate the best opportunities
in this area and have
identified possible signals worthy of further
study \cite{page95light,page95psi,cafe}.
In this letter we note that $e^+e^-$ annihilation in the vicinity of
charm threshold, 4 - 4.5 GeV in $E_{c.m.}$, offers special opportunities:
the lowest hybrid charmonia ``$H_{c}$'' are predicted
to exist at 4.1 - 4.2 GeV \cite{swanson94,perantonis90},
just above charm threshold where the conventional $c\bar{c}$ are rather well
understood and where there is a well known experimental anomaly which may
be due to excitation of these hybrid states \cite{ono84}.
As a result of our recent investigations
into the dynamics of the quark - flux--tube
system \cite{page95light,page95psi}
 we present new arguments that support this suggestion and propose
further experimental tests. In particular we shall argue that $\psi(4040)$ and
$\psi(4160)$ are strong mixtures of $\psi(3S)$ and hybrid charmonium.

% Thus for
%hybrid charmonia that are above the $DD^{**}(0^+,1^+,2^+)$ threshold

%$$
%H_c \rightarrow \hspace{-1.1em} / \hspace{0.6em}
%D\bar{D}, DD^*, D^*D^* ; H_c \rightarrow DD^{**}
%$$
%The decay channels for the $H_c$ are therefore rather sensitively dependent
%upon the mass since the $DD^{**}$ threshold is tantalisingly in the
region of the
%predicted mass $m(H_c)$. If $m(H_c) \gapproxeq 4.3$ GeV, the dominant
%modes will be to $DD^{**}$ and calculations \cite{page95conf}
following those of ref. \cite{page95light} show
%that the width turns on rapidly with increasing $m(H_c)$ hence
%making them hard to isolate.  Conversely, the width is
%suppressed if
%$m(H_c)\lapproxeq$ 4.3 GeV.
% as can be seen in Table \ref{4040table1}.
%The dominant decays may hence be through
%mixing between $H_c$ and conventional $c\bar{c}$ excitations.
%

One of us has shown \cite{page95psi} that, within the conventional
charmonium picture, the hadronic
width and branching ratios of $\psi(4040)$ and the narrowness of the
$\psi(4415)$
constrain the $\psi(4040)$  to be the $3S$ (as opposed to 2D)
 and the $\psi(4415)$ to be 4S (or possibly 5S) states; this
independently supports the spectroscopic assignments of many
potential models.
The $\psi(4160)$ hadronic width is consistent with it being any of
$3S,4S$
or
$2D$ \cite{page95psi} and hence, by elimination, one is tempted to
deduce
the $2D$
or possibly the $4S$ \cite{chao95}
assignment for it. However, $\Gamma^{ee}(4160) \sim \Gamma^{ee}(4040)$
which is inconsistent with any of these pictures.
%and moreover are too large for the $\psi(4160)$ to be dominantly $2D$
%(for which the leptonic width is
%strongly suppressed);
The simplest explanation is that these states are roughly $50:50$
mixtures of $\psi(3S)$ and
an ``inert" state that is essentially decoupled from $e^+e^-$
as originally suggested by Ono \cite{ono84}.

The unresolved problem is what  ``inert" state the $\psi(3S)$ is mixing with:
dynamics within standard spectroscopy
appear unable to
generate such a mixing consistently.
Mixing between $\psi(3S)-\psi(2D)$ via a tensor force
appears to be inadequate given the known limited mixing
for the $2S-1D$ states $\psi(3685;3770)$ \cite{ono84,cornell,tornqvist84}.
Strong mixing between $c\bar{c}$ and $D\bar{D}$ coupled channels is
incompatible with the observed narrow widths
(of order tens of MeV) of the $\psi(4040),\psi(4160)$ \cite{ono84}.
The maximum theoretical
mixing of $\psi(3S)-\psi(2D)$ generated in the literature, to the best

of our knowledge, is some
10\% \cite{ono84,cornell} in amplitude.

However, a mixing of $50$\% can arise naturally if there is mass
degeneracy between two
``primitive" states. The $\psi(3S)$ and $\psi(2D)$ are not expected
to be degenerate enough for large
mixing.\footnote{ A {\it coulomb + linear} potential gives
\protect\cite{barnesthanks} a 2D--3S mass difference of
79 - 87 MeV for string tension $b = 0.18 \pm 0.02\; {GeV}^{2}$, a
charm quark mass $m_{c} = 1.5 \pm 0.3$ GeV and $\alpha_{S} = 0.4$.
The ranges of $b$ and $m_{c}$ are the maximal ones consistent with
spectroscopy. The only way of reducing the 2D--3S mass difference
significantly appears to be by reducing $\alpha_{S}$; nonetheless
it is still $\sim 30$ MeV even in
the extreme limit $\alpha_{S} \rightarrow 0$.} However, a new
candidate
for an ``inert"
state degenerate with $\psi(3S)$ has recently emerged.

Numerical solutions of flux--tube charmonium spectroscopy predict
that the lightest hybrid charmonium states are in the 4.1 - 4.2 GeV
region \cite{swanson94}, while QCD inspired potential models with long

range linear
behaviour uniformly predict that
$\psi (3S$) is also in the narrow range 4.10 to 4.12
GeV \cite{cornell,richardson,bt}. This near degeneracy is also supported by
lattice studies of $c\bar{c}$ which
predict that
$\psi(3S)$ and $H_c$ are within 30 MeV of each other
\cite{perantonis90}
and, again, in the 4.1 - 4.2 GeV region. Consequently it is
probable that, within their widths, there will be mass degeneracy of
$\psi(3S)$
and $H_c$, leading to strong mixings and splitting of the eigenvalues.
If such a degeneracy occurs one immediately expects that the
physical eigenstates
will tend to be
\beqn
\psi_\pm \simeq \frac{1}{\sqrt{2}} (\psi(3S)\pm H_c)
\eeqn
which we shall identify as $\psi_- \equiv \psi(4040)$ and $\psi_+ \equiv
\psi(4160)$.

The $H_c$ component in eq.(1) will be ``inert" for the following reason.
A characteristic feature of hybrid mesons is the
prediction that their decays to
ground state mesons are suppressed and that the dominant coupling
is to excited
states \cite{page95light,kokoski85,page95conf,lipkin}, in
particular $DD^{**}$
for which the threshold is $\sim 4.3$ GeV. Hence
 this pathway will be closed for
$\psi(4040)$ and
$\psi(4160)$ and
consequently their hadronic decays will be driven by the
$\psi (3S)$
component. Thus the $H_c$ component
immediately satisfies the ``inert" criterion for the additional
piece and is partly responsible for the relatively narrow widths of these
states.

Both the dominant
production in  $e^+e^-$ annihilation and the prominent hadron
decays will be
driven by the $\psi (3S)$ component, leading to intimate
relationships between the properties of the two eigenstates.
First, this naturally explains their leptonic widths
$$
\Gamma^{e^+e^-}(\psi_+) \simeq  \Gamma^{e^+e^-}(\psi_-) \simeq
\frac{1}{2} \Gamma^{e^+e^-} (\psi(3S))
$$
which is consistent with the fact that potential
models uniformly predict a value for $\Gamma^{ee}(3S)$ that is
essentially a factor of two
larger than the data in both cases.

We now consider the implications for hadronic decays
with particular reference to
our recent dynamical analyses \cite{page95light,page95psi,swanson94}.

 The remarkable branching ratios (with phase space removed)
\beqn
\plabel{4040ratios}
\psi (4040)\rightarrow DD: DD^*: D^*D^* \simeq 1:20:640
\eeqn
differ considerably from those expected for a simple S--wave charmonium
$(1:4:7)$ and D--wave $(1:1:4)$ \cite{dgg,close76,cg}. Within a charmonium
picture the favoured interpretation has been that these ratios arise
as a consequence of nodes in the radial wave function
of $\psi(3S)$ which suppress $DD$ and $DD^*$ due to the coincidence of
$p$
(4040$\rightarrow DD, DD^*$) lying near the Fourier transformed
node (Figure 1)
 \cite{page95psi,leyaouanc77}.
This hypothesis is quantitatively consistent with the model description of
other $c\bar{c}$ dynamics \cite{page95psi} and we shall therefore
adopt this as our point of departure.
We note a useful kinematical coincidence.
\beqn
\plabel{4040a}
\frac{p (4160\rightarrow DD^*)}{p (4040\rightarrow DD)} =
\frac{0.75}{0.77}
\eeqn \intereqnvspace
\beqn
\plabel{4040b}
\frac{p(4160\rightarrow D^*D^*)}{p (4040\rightarrow DD^*)} = \frac{0.54}
{0.57}
\eeqn
If the decay amplitudes are insensitive to these small changes in
momenta then
one would expect immediate correlations along the following lines.
If the $DD$ and $DD^*$ channels of $\psi(4040)$ are suppressed due
to the node
structure of the $(3S)$ wave function, and
if the hadronic decays of both resonances  $\psi(4040)$ and $\psi(4160)$
are due to a common $\psi (3S)$
component, then one infers that there will be a corresponding
suppression of $DD^*$  and $D^* D^*$ for the $\psi$ (4160).
After taking due account of the spin weightings, one expects
\[
\Gamma(4160 \rightarrow DD^*) \simeq 4\Gamma(4040 \rightarrow DD)
\]

\intereqnvspace
\beqn
\plabel{4040naive}
\Gamma(4160 \rightarrow D^*D^*) \simeq \frac{7}{4}\Gamma(4040
\rightarrow DD^*)
\eeqn
from which one can deduce the na\"{\i}ve expectation
$\Gamma(4160 \rightarrow D^*D) : \Gamma(4160 \rightarrow D^*D^*)$ =
$1 : {3.5}^{+5.3}_{-1.3}$ using the ratios in ref. \cite{pdg94}.
Detailed calculations following the parameters, conventions and
methods of ref. \cite{page95psi} show modifications to these numbers
indicated in Table \ref{4040table2}.

\begin{table}[tb]
\begin{center}
\caption{\protect\small Width ratios in the flux--tube model
\protect\cite{paton85}
in regions A, B and C of parameter space. The various width ratios are
calculated at each ``dot'' in Figure \protect\ref{4040figure2} and the
mean $R$ and standard deviation $\sigma$ are computed for each
region.
The ``dots'' in
Figure \protect\ref{4040figure2} fit the experimental ratios
in eqn. \protect\ref{4040ratios} up to 1$\sigma$ deviations given in
ref. \protect\cite{pdg94}. This translates into $\sigma /R \sim
40\%$.
Ratios
indicated with $\sigma/R$ significantly smaller than this should hence
be understood to contain parameter independent information, and are hence
more model independent. No assumption about total widths are made.
The first ratio would still be valid for $\psi(4040)$ as pure
3S. The decays $\psi_{-} \rightarrow \protect\D \protect\D$ in Region B and
$\psi_{+} \rightarrow \protect\Ds \protect\D$ in Region C are at a
node and hence very
sensitive to parameters, indicated by an asterisk.
The conventions and parameters are those of ref. \protect\cite{page95psi}.}
\plabel{4040table2}
\begin{tabular}{|ll||c|r||c|r||c|r|}
\hline %------------------------
\multicolumn{2}{|c||}{Width Ratio} &  \multicolumn{2}{c||}{Region A}
&
 \multicolumn{2}{|c||}{Region B} & \multicolumn{2}{|c|}{Region C}\\
&&  $R$ &$\sigma/R$ & $R$ &$\sigma/R$ & $R$ &$\sigma/R$ \\
\hline \hline %------------------------
$\psi_{-} \rightarrow \ds \ds$ &/ $\psi_{-} \rightarrow \D \D$ &  4.0
&
   20\% &$\ast$&  $\ast$&   54 &   42\% \\
\hline \hline %------------------------
$\psi_{+} \rightarrow \ds \ds$ &/ $\psi_{+} \rightarrow \D \D$ &   36

   24\% &  5.8 &   13\% &   29 &   29\% \\
$\psi_{+} \rightarrow \ds \ds$ &/ $\psi_{+} \rightarrow \Ds\D$ &  1.4
&   30\% &  8.8 &   45\% &$\ast$&  $\ast$\\
$\psi_{+} \rightarrow D   D $  &/ $\psi_{+} \rightarrow \ds D$ &  3.3
&   37\% &  .25 &   58\% &  3.4 &   48\% \\
$\psi_{+} \rightarrow \ds \ds$ &/ $\psi_{+} \rightarrow D   D$ &  4.1
&   37\% &   10 &   30\% &  2.9 &    8\% \\
\hline \hline %------------------------
$\psi_{+} \rightarrow \ds D  $ &/ $\psi_{-} \rightarrow D   D$ &  2.3
&    8\% &  5.6 &    8\% &  2.1 &   14\% \\
$\psi_{+} \rightarrow \ds \ds$ &/ $\psi_{-} \rightarrow \ds D$ &  2.4
&    3\% &  1.2 &    5\% &  1.8 &    5\% \\
$\psi_{+} \rightarrow \Ds \D $ &/ $\psi_{-} \rightarrow \D \D$ &  3.9
&    3\% &$\ast$&$\ast$  &$\ast$&  $\ast$\\  
\hline \hline %------------------------
$\psi_{+} \rightarrow \ds \ds$ &/ $\psi_{-} \rightarrow \ds\ds$&  1.3
&   18\% &  .69 &   26\% &  1.1 &   23\% \\
$\psi_{-} \rightarrow \ds D  $ &/ $\psi_{+} \rightarrow \D \D$ &   15
&   22\% &  5.0 &    9\% &   16 &   24\% \\
\hline %------------------------------
\end{tabular}
\end{center}
\end{table}

The positions of the nodes in momentum space depend on the spatial spread of
the wave functions which are summarised by a parameter $\beta$ - see
Figure \ref{4040figure1}.
Depending on the magnitude of $\beta$ one can find different bands
of solutions
in $\beta$ space that fit the branching ratios of $\psi(4040) \rightarrow DD:
DD^*:D^*D^*$ (see Figure \ref{4040figure2}).
Historically the solutions\footnote{
$\beta_{A}=\beta=0.44$ GeV \protect\cite{leyaouanc77} was fitted from
experiment. $\beta_{A}=0.485$ GeV, $\beta=0.39$ GeV
\protect\cite{bradley}
and
$\beta_{A}=0.46$ GeV, $\beta=0.42$ GeV \protect\cite{alcock} were
deduced from numerical wave functions consistent with spectroscopy.
We obtain with a {\it coulomb + linear} potential
that $\beta_{A}= 0.41 - 0.48$ GeV
for a string tension $b = 0.18 \pm 0.02\; {GeV}^{2}$, a
charm quark mass $m_{c} = 1.5 \pm 0.3$ GeV and $\alpha_{S} = 0.4 \pm
0.1$.
Restricting to $b = 0.18 \; {GeV}^{2}$, $\alpha_{S} = 0.4$ and
$m_{c} = 1.5 - 1.8$ GeV preferred by potential models
\protect\cite{cornell},
we predict $\beta_{A}= 0.45 - 0.48$ GeV.
$(\beta_{A},\beta)$ is defined in Figure \protect\ref{4040figure2}.
}
of refs. \cite{leyaouanc77,bradley,alcock} appear to
be those of band ``A" where the outgoing momenta
$p_{D^*D}$ and $p_{DD}$ are both in the vicinity
of the first node, thereby suppressing the $DD$ and $DD^*$ channels.
A similar situation ensues for band ``C'' where both momenta are in
the vicinity of the second node.
Solution A gives
% a different pattern as small changes in $p$ near the
% node causes rapid fluctuation in relative rates***
$$
%\sigma (e^+e^- \rightarrow DD:DD^*:D^*D^*) \sim 2 to 7: 1 : 10
\sigma(e^+e^- \rightarrow DD:DD^*:D^*D^*:\D \D :\D \Ds) \sim
3 : 1 : 14 : 0.4 : 10
$$
Solution C gives qualitatively similar results to solution A (except
for the $\D \Ds$ mode)
$$
\sigma(e^+e^- \rightarrow DD:DD^*:D^*D^*:\D \D :\D \Ds) \sim
3 : 1 : 10 : 0.3 : 0
$$

In solution B,
$p_{DD}$ is near the second node and hence dramatically suppressed but
$p_{DD^*}$ is between the first and second nodes with a similar magnitude,
but opposite sign, to that in solution A.
When the magnitude of $p$ is varied by the amounts in
eqs. \ref{4040a} - \ref{4040b}
the effect on solution B is to move $DD^*$ into the region of a local maximum
and to shift the $DD$ away from the node (and hence to increase the
amplitude)
such that the effects of the radial nodes are washed out; as a result
the
rates
$DD:DD^*:D^*D^*$ tend to restore towards their ``naive" spin counting value
of 1:4:7. The actual ratios depend upon the parameters chosen within the band
(see Table \ref{4040table2}) the most probable results being
for solution B
$$
%\sigma(e^+e^- \rightarrow DD:DD^*:D^*D^*) \sim \frac{1}{2} to 3: 5 :8
\sigma(e^+e^- \rightarrow DD:DD^*:D^*D^*:\D \D :\D \Ds) \sim
1 : 4 : 10 : 2 : 1
$$

We shall consider the consequences of each of these ``mathematical''
solutions, even though the parameters of solution A are physically preferred.

The production of $D_s$ and $D_s^*$ from the $\psi(4040)$ and $\psi(4160)$
provide further consistency checks. The solution A which gave a dramatic
set of branching ratios for the $D$ system implies an effective ``spin
counting" result for the $D_s$ states arising from the common 3S
component of the $\psi(4040)$ and $\psi(4160)$. This is because
$$
p(\psi(4160) \rightarrow D_sD_s^*) = 0.41 \; GeV
$$
$$
p(\psi(4040) \rightarrow D_sD_s) = 0.45 \;  GeV
$$
and hence their
momenta are far from the nodes of the $(3S)$ component wave function.
On the other hand,
$$
p(\psi(4160) \rightarrow D_sD_s) = 0.67  \; GeV
$$
is coincident with the node: so in this scenario one has
the result that, due to the common $3S$ component, the spin counting is
realised best between the two states
$$
\Gamma(\psi(4160) \rightarrow D_sD_s^*) \sim 4 \Gamma(\psi(4040)
 \rightarrow D_sD_s)
$$
while
$$
\Gamma(\psi(4160) \rightarrow D_sD_s) \sim 0.
$$

This characteristic set of results is very different from those in solution B
where proximity of the node suppresses $\Gamma(\psi(4040)
 \rightarrow D_sD_s) \sim 0$.

Thus in summary the three patterns that arise in this $\psi(3S) - H_c$ mixing
picture of the $\psi(4040)$ and $\psi(4160)$ are :

Solution A (physically realistic)
$$
\psi(4040) \rightarrow  DD<D_sD_s < D^*D < D^*D^*
$$
\intereqnvspace
\beqn
\plabel{4040rat1}
\psi(4160) \rightarrow D_sD_s \lapproxeq D^*D < DD \lapproxeq D_sD_s^*
\approx D^*D^*
\eeqn

Solution B
$$
\psi(4040) \rightarrow D_sD_s <DD < D^*D < D^*D^*
$$
\intereqnvspace
\beqn
\plabel{4040rat2}
\psi(4160) \rightarrow DD \approx D_s^*D_s \approx D_sD_s < D^*D  <  D^*D^*
\eeqn

Solution C
$$
\psi(4040) \rightarrow D_sD_s <DD < D^*D < D^*D^*
$$
\intereqnvspace
\beqn
\plabel{4040rat3}
\psi(4160) \rightarrow  D_s^*D_s <  D_sD_s \lapproxeq D^*D < DD  <  D^*D^*
\eeqn
with quantitative measures given in table 1.

Independent evidence for the $\psi(4040)$ and $\psi(4160)$
consisting of $\psi(3S)$ mixed with an inert state arises
within this model from a
study of their total widths. Figure 2 shows that
the regions in parameter space consistent with
the experimental widths are the same
independent of the pair creation amplitude, implying a common dynamics.

If the tests in eqs. \ref{4040rat1} - \ref{4040rat3} are successful
they will confirm the common presence of the
$(3S)$
component in the wave functions and that the additional component is
inert to
hadronic decays thereby implying a new dynamics beyond that in conventional
charmonium. Proving that this is due to hybrid states would then
require further
signals in other partial waves since the above mixing will ironically
have made the
hybrid dynamics effectively invisible in the $1^{--}$ channel. For example,
vector hybrids would not then be a significant
source of enhanced $\psi(3685)$ at CDF \cite{close95}.
 However, if the exotic
$1^{-+}$ state lies below the $\psi(4040)$ or $\psi(4160)$, there is
the
possibility
of a direct spin flip M1 radiative transition of the hybrid component
$H_c \rightarrow \gamma 1^{-+}$
exposing the exotic hybrid charmonium unambiguously. The matrix
element for $H_c \rightarrow \gamma 1^{-+}$ is related to that of
$\psi \rightarrow \eta_c \gamma$ in the limit where
the photon momentum tends to zero. The M1 transition involves
$c\bar{c}$
spin
flip from $S=1$ to $S=0$, the only additional feature involving the coupling
of spin and the orbital angular momentum associated with the excited
flux--tube
for the $H_c = 1^{-+}$ state. To the extent that the flux--tube excitation is
transversely polarised \cite{page95light,paton85} it is
effectively $L_z=\pm 1$. Hence,

$$
M=\langle\psi_{\pm}|\vec{\mu}|H_c(1^{-+})\rangle = cos\theta \;
\langle 11,10|11\rangle \; \mu
$$
where $\mu$ is the strength of the M1 transition $\psi \rightarrow \eta_c
 \gamma$ and $\cos\theta
\sim \frac{1}{\sqrt{2}}$ is the $\langle \psi_{\pm}|H_c \rangle$
mixing angle.
 Thus the relative rates scale as
$$
\Gamma(\psi_{\pm} \rightarrow \gamma H_c(1^{-+})) \sim 0.3 \;
(\frac{p}{118 \; MeV})^3 \; keV
$$
implying a branching ratio of $O(10^{-5})$. The subsequent transition

$$
\psi_{\pm} \rightarrow H_c(1^{-+}) + \gamma \rightarrow
\psi(3095) \gamma \gamma
$$
may provide the $\psi(3095)$ as a tag,
though a dedicated search at a high intensity Tau Charm Factory may
still be required
to isolate this signal.

There may be analogous signals in the $\Upsilon$ spectrum. There is a
mass shift of the
$\Upsilon(4S)$ and the candidate $\Upsilon(5S)(10580)$ that
appears qualitatively
similar to those of $\psi(4040)$ and $\psi(4160)$
which can conceivably be explained by an increase of 80 MeV in the
$\Upsilon(5S)-\Upsilon(4S)$ mass separation due to coupled channel
effects  \cite{tornqvist84}. Futhermore, there is
 a small discrepancy between
leptonic widths and theory but not at any significant level. Any
mixing with an ``inert" hybrid here would require further states
 \cite{ono84} to be seen, albeit at a low level,
since the highest mass state $\Upsilon(11020)$ is the sixth in the tower and
is consistent with being $6S$, hence militating against strong mixings with
additional degrees of freedom. Given that there will soon be
extensive studies
in the $4S$ peak at B-factories it may be interesting to see if there are
any anomalous radiative decays to $1^{-+}$ $H_b$ state arising
from a possible
$H_b$ component mixed into the $4S$ peak. However, we consider the charmonium
states $\psi(4040)-\psi(4160)$ currently to be the most likely
examples of hybrid
mixing in the heavy quark sector and we urge a dedicated study
of the $\psi(4160)$
in particular to test this hypothesis.

\vspace{0.3cm}
We wish to thank T. Barnes, J.M. Richard and W. Toki for comments.

\begin{figure}
\vspace{-2.5cm}
\centerline{ \psfig {figure=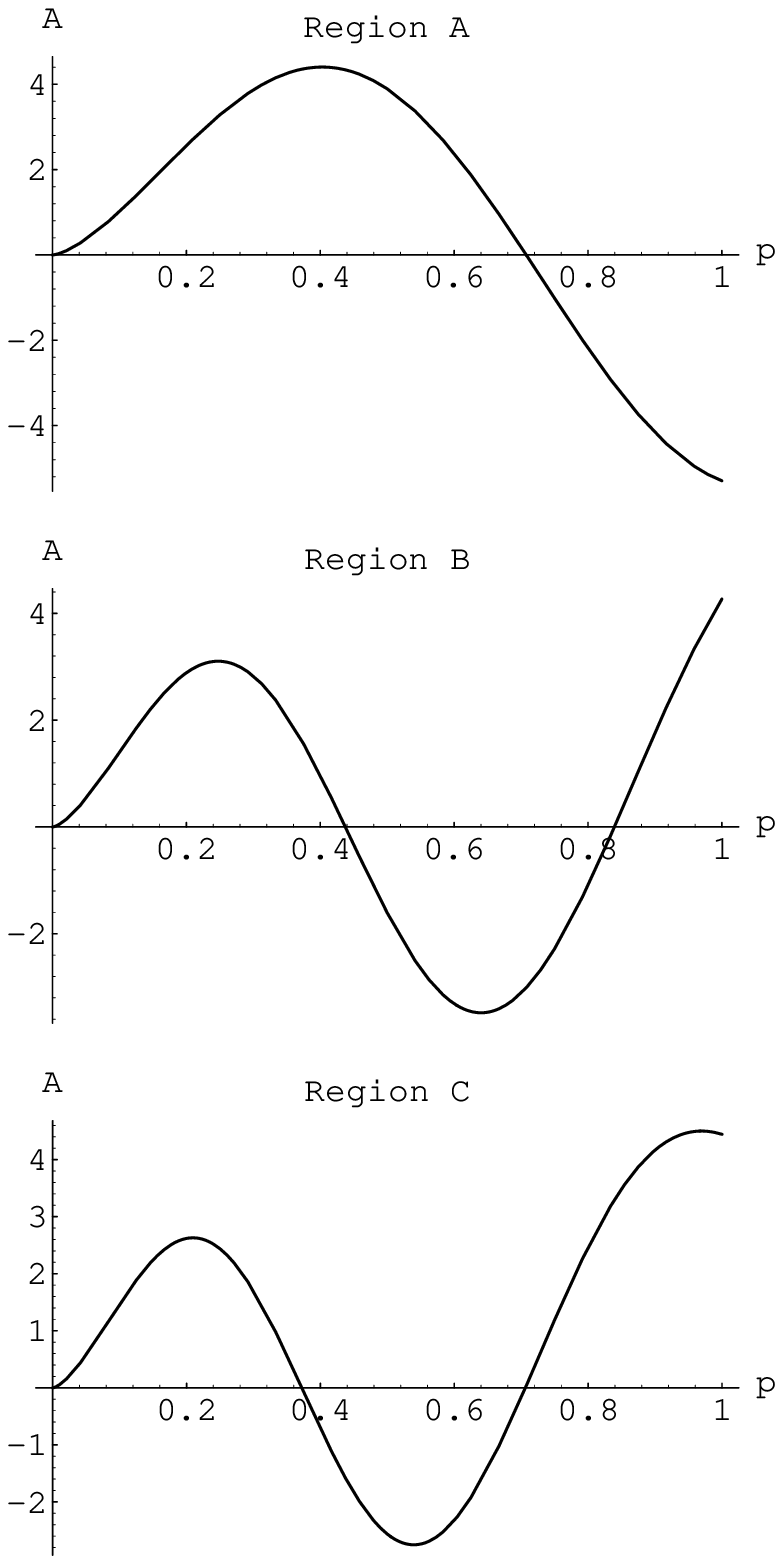,height=20cm,width=20cm}}
\vspace{-2.5cm}
\caption{\protect\small The $3S \rightarrow 1S + 1S$ amplitude
$A \equiv \pm \protect\sqrt{\tilde{\Gamma}(c\bar{c} \rightarrow
c\bar{u}+\bar{c}u)}$
in $MeV^{1/2}$ as a function of the CM momentum $p$ in GeV.
The physical $\Gamma = SF \: \; \tilde{\Gamma} \; \tilde{M_{B}}
\tilde{M_{C}} /
\tilde{M_{A}}
\simeq SF \: \; \tilde{\Gamma}$, where the spin--averaged masses
$\tilde{M}$
are defined in ref. \protect\cite{page95psi}, and the spin--flavour factor
$SF$ is 1 ($DD$), 4 ($\protect\ds D$), 7 ($\protect\ds \protect\ds$),
$\frac{1}{2}$ ($\protect\D \protect\D$) or 2 ($\protect\Ds \protect\D$).
We display $A$ at typical $(\beta_{A},\beta)$ (in GeV) :
$(0.475,0.455)$ (Region A), $(0.290,0.265)$ (Region B) and
$(0.245,0.215)$ (Region C).
The amplitude is slightly modified for
$\tilde{\Gamma}(c\bar{c} \rightarrow c\bar{s}+\bar{c}s)$ due to a
$s$--quark mass differing from the $u$--quark \protect\cite{page95psi}.}
\plabel{4040figure1}
\end{figure}

\begin{figure}
\centerline{\psfig {figure=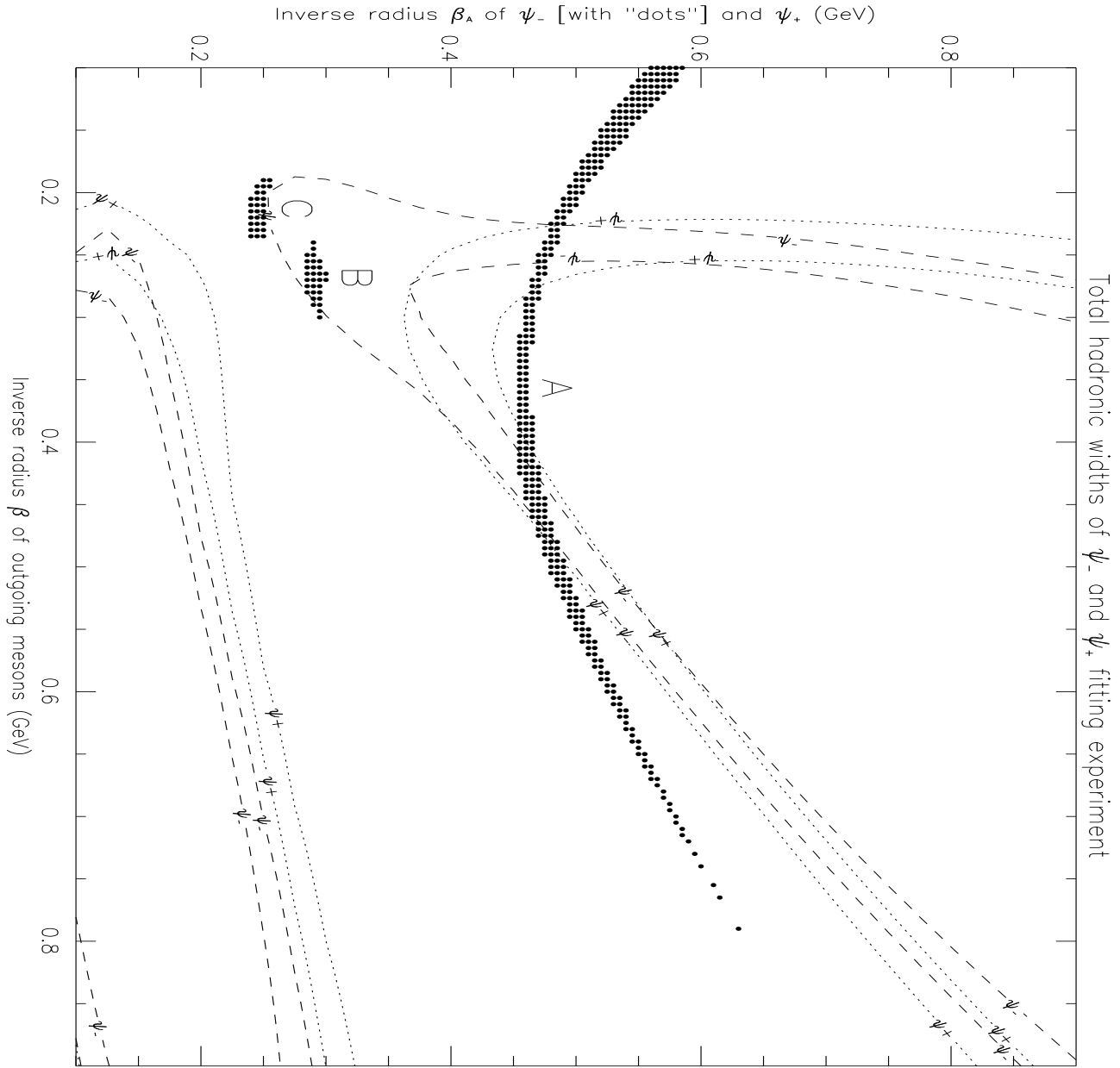,height=17.5cm,width=17.5cm}}
\caption{\protect\small Total hadronic widths of
$\psi_{-} \equiv \psi(4040)$ and $\psi_{+} \equiv \psi(4160)$ fitting
experiment. The contours enclose narrow regions with
experimentally acceptable \protect\cite{pdg94}
total widths, indicating 1$\sigma$ variations from
the mean experimental width. The regions where the $\psi_{-}$
decay ratios fit
experiment are indicated by the ``dots''. We denote them by Regions A,
B and C.}
\plabel{4040figure2}
\end{figure}

\end{document}